\documentclass[conference]{IEEEtran}
\IEEEoverridecommandlockouts

\usepackage[T1]{fontenc}
\usepackage[utf8]{inputenc}
\usepackage{csquotes}

\usepackage{amsfonts}
\usepackage{amsmath}
\usepackage{braket}
\usepackage{pifont}
\usepackage{nicefrac}

\usepackage{graphicx}
\usepackage{tikz}
\usetikzlibrary{quantikz2}
\usetikzlibrary{arrows.meta}

\usepackage{tabularray}
\UseTblrLibrary{booktabs}

\usepackage{parskip}
\usepackage{subcaption}

\usepackage{xcolor}
\usepackage{orcidlink}
\usepackage{hyperref}
\hypersetup{colorlinks=true,urlcolor=teal,
  linkcolor=teal,citecolor=teal}
\usepackage{glossaries-extra}

\usepackage[export]{adjustbox}
\usepackage[inline]{enumitem}
\usepackage{xspace}
\usepackage{censor}

\usepackage{microtype}

\newif\ifshowtodos
\showtodosfalse
\ifshowtodos\usepackage{todonotes}\else\usepackage[disable]{todonotes}\fi

\newboolean{anonymous}
\setboolean{anonymous}{false}

\usepackage[
  backend=biber,
  safeinputenc=true, 
  maxcitenames=3,
  minbibnames=3,
  mincitenames=1,
  maxbibnames=3,
  uniquelist=false,
  url=false,
  doi=true,
  eprint=false,
  isbn=false,
]{biblatex}
\setabbreviationstyle[acronym]{long-short}
\glssetcategoryattribute{acronym}{nohyperfirst}{true}

\newacronym{fcc}{FCC}{Fourier coefficient correlation}
\newacronym{fft}{FFT}{fast Fourier transform}
\newacronym{fm}{FM}{feature map}
\newacronym{ftqc}{FTQC}{fault‑tolerant quantum computing}
\newacronym{hea}{HEA}{hardware‑efficient ansatz}
\newacronym{hep}{HEP}{high‑energy physics}
\newacronym{hs}{HS}{Hilbert-Schmidt}
\newacronym{kl}{KL}{Kullback‑Leibler}
\newacronym{lhc}{LHC}{Large Hadron Collider}
\newacronym{lo}{LO}{leading order}
\newacronym{ml}{ML}{machine learning}
\newacronym{mlp}{MLP}{multi‑layer perceptron}
\newacronym{mse}{MSE}{mean squared error}
\newacronym{nisq}{NISQ}{noisy intermediate‑scale quantum}
\newacronym{nlo}{NLO}{next‑to‑leading order}
\newacronym{pca}{PCA}{principal component analysis}
\newacronym{pcc}{PCC}{pearson correlation coefficient}
\newacronym{pqc}{PQC}{parameterized quantum circuit}
\newacronym{qaoa}{QAOA}{quantum approximate optimisation algorithm}
\newacronym{qc}{QC}{quantum computing}
\newacronym{bp}{BP}{barren plateau}
\newacronym{qcd}{QCD}{quantum chromodynamics}
\newacronym{qfm}{QFM}{quantum Fourier model}
\newacronym{qk}{QK}{quantum kernel}
\newacronym{qml}{QML}{quantum machine learning}
\newacronym{qnn}{QNN}{quantum neural network}
\newacronym{rff}{RFF}{random Fourier features}
\newacronym{spam}{SPAM}{state preparation and measurement}
\newacronym{qoc}{QOC}{quantum optimal control}
\newacronym{qpu}{QPU}{quantum processing unit}
\newacronym{qft}{QFT}{quantum Fourier transform}
\newacronym{dla}{DLA}{dynamical Lie algebra}
\newacronym{lasa}{LASA}{Lie algebra supported ansatz}
\newacronym{mga}{MGA}{matchgate ansatz}
\newacronym{sea}{SEA}{strongly entangling ansatz}
\newacronym{la}{LA}{Lie-algebraic}

\newcommand{\ie}{\emph{i.e.}\xspace}
\newcommand{\wrt}{\emph{w.r.t.}\xspace}
\newcommand{\cf}{\emph{cf.}\xspace}

\newcommand{\ase}{ansatzes\xspace}

\newcommand{\as}{ansatz\xspace}

\newcommand{\btheta}{\ensuremath \boldsymbol{\theta}}
\newcommand{\bphi}{\ensuremath \boldsymbol{\phi}}
\newcommand{\bw}{\ensuremath \boldsymbol{w}}
\newcommand{\bepsilon}{\ensuremath \boldsymbol{\epsilon}}
\newcommand{\bx}{\ensuremath \boldsymbol{x}}
\newcommand{\expphi}{\ensuremath \mathbb{E}_{\bphi}}
\newcommand{\loss}{\ensuremath \mathcal{L}}
\newcommand{\obs}{\ensuremath O}
\newcommand{\pur}{\ensuremath \mathcal{P}}
\newcommand{\var}{\ensuremath \operatorname{Var}}
\newcommand{\mg}{\mathfrak{g}}
\newcommand{\mso}{\mathfrak{so}}
\newcommand{\god}{{\mathfrak{g}_{\mathrm{od}}}}
\newcommand{\gh}{{\mathfrak{g}_{\mathrm{h}}}}

\newcommand{\imag}{\mathrm{i}}

\DeclareMathOperator{\tr}{Tr}
\newcommand{\ketbra}[2]{\ket{#1}\!\bra{#2}}

\newcommand{\RY}{\ensuremath{R_y}}
\newcommand{\RZ}{\ensuremath{R_z}}
\newcommand{\iid}{i.i.d.\xspace}

\definecolor{color1}{HTML}{009371}
\definecolor{color2}{HTML}{E69F00}
\definecolor{color3}{HTML}{1f78b4}
\definecolor{color4}{HTML}{002D4C}

\ifbool{anonymous}{}{\renewcommand{\blackout}[1]{#1}}
\ifbool{anonymous}{\newcommand{\geneurl}[1]{\href{https://thiscatexists.com/}{\blackout{#1}}}}{\newcommand{\geneurl}[1]{\url{#1}}}
\ifbool{anonymous}{\renewcommand{\orcidlink}[1]{}}{}

\begin{document}
\title{Unflattening by Flattening --\\
  How Input Distributions Shape Output Variance in Angle-Encoded Circuits}

\author{
  \IEEEauthorblockN{%
    \blackout{Melvin Strobl}\orcidlink{0000-0003-0229-9897},
    \blackout{Gabriel Mejia}\orcidlink{0009-0009-8002-1226},
    \blackout{Eileen Kuehn}\orcidlink{0000-0002-8034-8837},
    \blackout{Achim Streit}\orcidlink{0000-0002-5065-469X}
  }

  \IEEEauthorblockA{
    Karlsruhe Institute of Technology, Germany
  }
}

\maketitle{}

\begin{abstract}
  \unboldmath 
  Barren plateaus hinder training of parameterized quantum circuits by making loss gradients exponentially small in the number of qubits.
  For angle-encoded product states, we show how the input distribution affects output variation through algebraic input purity, \ie the input's overlap with the circuit's \glsentrylong{dla}.
  With the specified readouts and Haar or exact group 2-design sampling, matchgate circuits on $n$ qubits retain output variance of order $1/n$ for every pure product input.
  The off-diagonal family instead has zero output on computational-basis inputs at any depth and for every parameter choice.
  Independent uniform angles yield mean variance of the same order as matchgates.
  A count of diagonal Pauli strings identifies these zero-output endpoints.
  For any fixed dataset of nonzero inputs with at least $2n$ coordinates, we construct a classical preprocessing certificate.
  A shared random rotation is accepted only when the dataset's mean purity passes a computable threshold.
  This certifies mean output variance of order $1/n$ for the off-diagonal family before circuit execution, with a constant expected number of trials.
  Binary and ternary weighted encodings also recover the independent-angle mean purity from one uniform scalar input.
  Numerical experiments examine training behavior and extensions to larger algebras.
  The guarantee concerns output variance averaged over inputs, while successful learning also depends on the task and information preserved by the encoding.
\end{abstract}


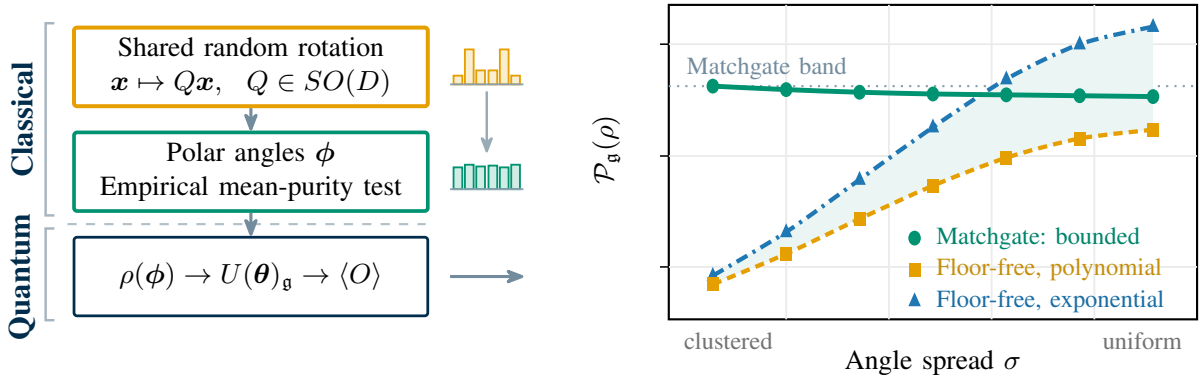
\begin{figure*}[!htb]
  \centering
  \resizebox{0.875\textwidth}{!}{%
\begin{tikzpicture}[
    font=\small,
    >={Stealth[round]},
    lbox/.style={rounded corners=1pt, align=center, inner sep=3pt,
                 minimum width=4.05cm, minimum height=0.9cm},
    dnode/.style={lbox, draw=color4!55, line width=0.9pt, fill=white},
    pcbox/.style={lbox, draw=color2, line width=1.3pt, fill=none},
    pnode/.style={lbox, draw=color1, line width=1.2pt, fill=none},
    qnode/.style={lbox, draw=color4, line width=0.9pt, fill=none},
]

\node[rotate=90, color4, font=\bfseries] at (0.18,2.6) {Classical};
\node[rotate=90, color4, font=\bfseries] at (0.18,0.8) {Quantum};
\draw[color4!35, line width=0.8pt] (0.55,1.5)--(0.45,1.5)--(0.45,3.7)--(0.55,3.7);
\draw[color4!35, line width=0.8pt] (0.55,0.3)--(0.45,0.3)--(0.45,1.35)--(0.55,1.35);
\draw[color4!30, dashed, line width=0.7pt] (0.7,1.4)--(4.8,1.4);

\node[pcbox] (pc)    at (2.8,3.2) {Shared random rotation\\[1pt]
       $\bx\mapsto Q\bx,\ \ Q\in SO(D)$};
\node[pnode] (prior) at (2.8,2.0) {Polar angles $\bphi$\\[1pt]Empirical mean-purity test};
\node[qnode] (qm)    at (2.8,0.8)  {$\rho(\bphi)\to U(\btheta)_{\mg}\to\langle\obs\rangle$};

\draw[->, color4!55, line width=1.0pt] (pc) -- (prior);
\draw[->, color4!55, line width=1.0pt] (prior) -- (qm);

\begin{scope}[shift={(5.5,3.0)}]
  \foreach \bx/\bh in {-0.33/0.10,-0.20/0.40,-0.07/0.16,0.07/0.16,0.20/0.40,0.33/0.10} {
    \fill[color2!18] (\bx-0.05,0) rectangle (\bx+0.05,\bh);
    \draw[color2, line width=0.6pt] (\bx-0.05,0) rectangle (\bx+0.05,\bh);
  }
  \draw[color4!45, line width=0.6pt] (-0.42,0) -- (0.42,0);
\end{scope}

\begin{scope}[shift={(5.5,1.8)}]
  \foreach \bx/\bh in {-0.33/0.25,-0.20/0.28,-0.07/0.26,0.07/0.27,0.20/0.25,0.33/0.28} {
    \fill[color1!18] (\bx-0.05,0) rectangle (\bx+0.05,\bh);
    \draw[color1, line width=0.6pt] (\bx-0.05,0) rectangle (\bx+0.05,\bh);
  }
  \draw[color4!45, line width=0.6pt] (-0.42,0) -- (0.42,0);
\end{scope}

\draw[->, color4!40, line width=0.8pt] (5.5,2.92) -- (5.5,2.22);
\draw[->, color4!55, line width=1.0pt] (5.08,0.8) -- (5.92,0.8);

\begin{scope}[shift={(5.4,-0.24)}, xscale=0.84]
  \draw[black, line width=0.7pt] (2.6,0.55) rectangle (9.8,4.15);
  \node[black] at (6.2,0.06) {Angle spread $\sigma$};
  \node[black!55, font=\footnotesize, anchor=west] at (2.66,0.30) {clustered};
  \node[black!55, font=\footnotesize, anchor=east] at (9.74,0.30) {uniform};
  \node[black, rotate=90] at (1.75,2.5) {$\pur_{\mg}(\rho)$};
  \foreach \y in {3.7,2.42,1.15} {
    \draw[black!8, line width=0.6pt] (2.6,\y) -- (9.8,\y);
    \draw[black, line width=0.5pt] (2.6,\y) -- (2.48,\y);
  }
  \foreach \x in {4.2,5.6,7.0,8.4}
    \draw[black!8, line width=0.6pt] (\x,0.55) -- (\x,4.15);
  \node[color4!55, font=\footnotesize, anchor=west] at (2.70,3.42) {Matchgate band};
  \draw[color4!45, line width=0.7pt, dotted] (2.6,3.22) -- (9.8,3.22);
  \fill[color1!8]
    (3.2,1.05) -- (4.2,1.55) -- (5.2,2.15) -- (6.2,2.75) -- (7.2,3.30)
    -- (8.2,3.70) -- (9.2,3.90) -- (9.2,2.72) -- (8.2,2.62) -- (7.2,2.40)
    -- (6.2,2.08) -- (5.2,1.70) -- (4.2,1.30) -- (3.2,0.95) -- cycle;
  \draw[color3, line width=1.4pt, dash pattern=on 4pt off 2pt on 1pt off 2pt]
    plot[smooth] coordinates
    {(3.2,1.05)(4.2,1.55)(5.2,2.15)(6.2,2.75)(7.2,3.30)(8.2,3.70)(9.2,3.90)};
  \foreach \x/\y in {3.2/1.05,4.2/1.55,5.2/2.15,6.2/2.75,7.2/3.30,8.2/3.70,9.2/3.90}
    \fill[color3] (\x,\y) ++(-0.085,-0.06) -- ++(0.17,0) -- ++(-0.085,0.16) -- cycle;
  \draw[color2, line width=1.4pt, densely dashed] plot[smooth] coordinates
    {(3.2,0.95)(4.2,1.30)(5.2,1.70)(6.2,2.08)(7.2,2.40)(8.2,2.62)(9.2,2.72)};
  \foreach \x/\y in {3.2/0.95,4.2/1.30,5.2/1.70,6.2/2.08,7.2/2.40,8.2/2.62,9.2/2.72}
    \fill[color2] (\x,\y) +(-0.075,-0.075) rectangle +(0.075,0.075);
  \draw[color1, line width=1.6pt] plot[smooth] coordinates
    {(3.2,3.22)(4.2,3.18)(5.2,3.15)(6.2,3.13)(7.2,3.12)(8.2,3.11)(9.2,3.10)};
  \foreach \x/\y in {3.2/3.22,4.2/3.18,5.2/3.15,6.2/3.13,7.2/3.12,8.2/3.11,9.2/3.10}
    \fill[color1] (\x,\y) circle (0.085);
  \fill[color1] (5.95,1.48) circle (0.085);
  \node[color1!85!color4, font=\footnotesize, anchor=west] at (6.10,1.48)
    {Matchgate: bounded};
  \fill[color2] (5.95,1.12) +(-0.075,-0.075) rectangle +(0.075,0.075);
  \node[color2!85!color4, font=\footnotesize, anchor=west] at (6.10,1.12)
    {Floor-free, polynomial};
  \fill[color3] (5.95,0.76) ++(-0.085,-0.06) -- ++(0.17,0) -- ++(-0.085,0.16) -- cycle;
  \node[color3!85!color4, font=\footnotesize, anchor=west] at (6.10,0.76)
    {Floor-free, exponential};
\end{scope}

\end{tikzpicture}%
}
  \caption{\emph{Left:} A shared random rotation and polar-angle extraction produce encoding angles.
    A classical acceptance test checks their empirical mean off-diagonal purity before the quantum circuit is run.
    \emph{Right:} Schematic purity responses to the angle spread, without a numerical scale.
    The matchgate family (solid) stays in a narrow band while the floor-free examples (dashed and dash-dotted) approach zero at computational-basis inputs.
    Their output variances additionally depend on the simple-component dimensions and observable projections in~\autoref{eq:variance}.}
  \label{fig:overview}
\end{figure*}

\section{Introduction}
\label{sec:introduction}

\Glspl{bp} are a central obstacle to training \glspl{pqc} at scale in \gls{qml}.
Several ways of diagnosing them begin with a basic question: how much does the model output change when the circuit parameters vary?
For an $n$-qubit input state $\rho$ evolved by $U(\btheta)$, denote the measured output by $f_{\btheta}(\rho,\obs)=\tr[U(\btheta)\rho\,U^\dagger(\btheta)\obs]$.
Its variance over a chosen parameter ensemble measures the spread of outputs for this fixed input.
If $\var_{\btheta}[f_{\btheta}]\in\mathcal{O}(b^{-n})$ for some $b>1$, the output landscape concentrates exponentially under the chosen parameter ensemble.
Connecting this concentration to exponentially small gradients requires further assumptions~\cite{mcclean_barren_2018,cerezo_cost_2021,arrasmith_equivalence_2022,ragone_lie_2024}.
Throughout this work, we therefore distinguish sample-wise output variance from supervised training loss $\loss$ and its gradients.

In the noiseless setting, the circuit's \gls{dla} $\mg=\langle \imag\mathcal{G}\rangle_{\mathrm{Lie}}$ is the Lie closure of its gate generators~\cite{larocca_diagnosing_2022,ragone_lie_2024}.
Concretely, one starts with the operators generating the available gates and includes their commutators until no new independent operators appear.
This linear space describes the infinitesimal transformations accessible through those gates.
When the readout lies in this algebra, $\obs\in \imag\mg$, conjugating it by the circuit mixes operators within the same space.
This is the setting of a \gls{lasa}~\cite{fontana_characterizing_2024}.

The input's $\mg$-purity measures its overlap with that operator space~\cite{barnum_generalizations_2003,somma_quantum_2005,ragone_lie_2024} which is the sum of squared expectations of the algebra's basis operators.
It therefore differs from ordinary state purity $\tr[\rho^2]$, which equals one for every pure state.
All inputs in our baseline analysis are pure product states, yet their $\mg$-purity can vary because their overlap with the circuit's operators changes with the encoding angles.
If that overlap vanishes, every in-algebra readout has zero expectation throughout the circuit's evolution.

The variance formula combines these input and readout overlaps with the algebra dimension.
As a subalgebra of $\mathfrak{u}(2^n)$, the \gls{dla} is reductive and decomposes into commuting components $\mg=\mg_1\oplus\dots\oplus\mg_{k-1}\oplus\mg_k$, where the $\mg_j$ are simple Lie algebras for $j<k$ and $\mg_k$ is the abelian center~\cite{ragone_lie_2024}.
Each simple component is an operator sector that the circuit mixes internally, while operators in the center remain unchanged.
Haar sampling draws uniformly from the transformations in the dynamical Lie group $e^{\mg}$.
An exact group 2-design likewise reproduces the moments needed to compute output mean and variance.
For Haar sampling on the dynamical Lie group $e^{\mg}$, or an exact group 2-design, the \gls{la} theory gives the output variance \wrt the parameters (Theorem~1 of~\cite{ragone_lie_2024}, \cf~\cite{fontana_characterizing_2024}):
\begin{equation}
  \var_{\btheta}\left[f_{\btheta}(\rho,\obs)\right] \;=\; \sum_{j=1}^{k-1} \frac{\pur^{\mathrm{HS}}_{\mg_j}(\rho)\, \pur^{\mathrm{HS}}_{\mg_j}(\obs)}{\dim \mg_j}.
  \label{eq:variance}
\end{equation}

Each term pairs the input and readout projections within one simple component and divides by its dimension.
Thus, a component can contribute to output variation only if both projections are nonzero.
By contrast, the abelian center contributes only to the output mean~\cite{ragone_lie_2024}.
The \gls{hs} $\mg$-purity $\pur^{\mathrm{HS}}_\mg(H)=\sum_{\alpha}\tr[B_\alpha H]^2$ measures the squared projection of a Hermitian operator $H$ onto $\imag\mg$.
Here $\{B_\alpha\}$ is any \gls{hs}-orthonormal basis of $\imag\mg$, satisfying $\tr[B_\alpha B_\beta]=\delta_{\alpha\beta}$ (Eq.~(7) of~\cite{ragone_lie_2024}, \cf~\cite{barnum_generalizations_2003,somma_quantum_2005}).

For the families studied here, bare Pauli strings $P$ form an orthogonal basis $\mathcal{B}_\mg$ of $\imag\mg$ with $\tr[P^2]=2^n$.
This follows from their Pauli-string generators, whose nested commutators remain proportional to Pauli strings.
We use the corresponding Pauli-basis purity $\pur_\mg(\rho)=\sum_{P\in\mathcal{B}_\mg}\braket{P}^2_{\rho}=2^n\pur^{\mathrm{HS}}_\mg(\rho)$ in the closed forms below.
For algebras without such a basis, \autoref{eq:variance} remains valid in its \gls{hs} form, but the string-counting formulas do not apply.
For a Pauli-string observable $\obs\in \imag\mg$, these normalization factors cancel, since $\pur^{\mathrm{HS}}_\mg(\obs)=\tr[\obs^2]=2^n$.
Thus, for a simple \gls{dla}, \autoref{eq:variance} reduces to $\var_{\btheta}[f_{\btheta}]=\pur_\mg(\rho)/\dim\mg$ (\cf \citeauthor{ragone_lie_2024}~\cite{ragone_lie_2024} and~\autoref{apx:purity}).
Because $\pur_\mg(\rho)$ is evaluated before the circuit acts, it is independent of $\btheta$.
The circuit rotates the input's projection within each component while preserving its size, so varying the trainable parameters cannot create a missing projection~\cite{ragone_lie_2024}.

For an approximate design, the variance formula has an error determined by the difference from Haar second moments~\cite{ragone_lie_2024}.
Available rapid-mixing results also depend on how generators are sampled, so they do not by themselves supply a mixing depth for every brickwork circuit considered here~\cite{fontana_characterizing_2024}.

\autoref{eq:variance} shows why polynomial \gls{dla} dimension alone does not rule out a \gls{bp}.
The variance can still be exponentially small when the encoded input has exponentially small $\mg$-purity, consistent with earlier conditions for avoiding \glspl{bp}~\cite{thanasilp_subtleties_2023,ragone_lie_2024,cerezo_does_2025}.
We make this input dependence explicit for angle-encoded product states and turn it into a testable preprocessing guarantee.
Our formulas distinguish a family whose output-variance scaling is protected for every pure product input from a family whose output vanishes at computational-basis inputs.
For the latter, a shared random rotation followed by polar-angle extraction can certify an inverse-polynomial empirical mean output variance under the specified circuit ensemble.
Previous studies have examined related data-dependent effects from different perspectives.
\citeauthor{thanasilp_subtleties_2023}~\cite{thanasilp_subtleties_2023} trace dataset-induced plateaus to the entanglement and mixedness of the encoded states.
In Ref.~\cite{barthe_gradients_2024}, \citeauthor{barthe_gradients_2024} bound the gap between the data-averaged gradient variance and its data-free counterpart.
\citeauthor{kulshrestha_exponentially_2026}~\cite{kulshrestha_exponentially_2026} study how the parameter-initialization distribution affects \glspl{bp}.
\citeauthor{ohno_observable_2026}~\cite{ohno_observable_2026} instead select circuit generators using the observable, with a purity interpretation under restricted settings.
In \citetitle{strobl_fourier_2026}~\cite{strobl_fourier_2026}, correlations between Fourier coefficients are used to assess data re-uploading models before training and predict relative \as performance on the tasks studied.
Our purity certificate provides a complementary diagnostic of output variation for a fixed circuit, readout, and dataset.

In this work we keep the circuit and readout fixed and ask how the induced angle distribution changes input purity.
Specifically, we
\begin{enumerate*}[label=(\roman*)]
  \item derive the matchgate purity band $[n-1,n]$ for every pure product input;
  \item identify clustered inputs with zero in-algebra output by counting diagonal Pauli basis strings;
  \item derive the off-diagonal purity formula and its uniform-prior mean, then construct a shared-rotation acceptance test for any fixed dataset satisfying the preprocessing assumptions;
  \item establish an exact mean-purity identity for dissociated scalar encodings and examine data re-uploading numerically; and
  \item test how clustering affects purity in families with exponential algebras.
\end{enumerate*}
To interpret this certificate, we compute output variance over circuit parameters for each input, then average these variances over the dataset.
We call this the mean sample-wise output variance.
It characterizes the output variation available on the encoded inputs, while the supervised loss also depends on their labels and on how their outputs vary together.
For the off-diagonal family, spreading the angles uniformly can raise this mean variance, giving a precise sense in which flattening the angle distribution can unflatten the output landscape.
\autoref{fig:overview} sketches the setup and the purity responses that motivate the analysis.
We specify the preprocessing, encoding, and circuit families in~\autoref{sec:background}, then develop the closed forms and their numerical checks in~\autoref{sec:results}.
Finally, we discuss the main limitations and practical interpretation in~\autoref{sec:outlook}.

\section{Background}
\label{sec:background}

\paragraph{Preconditioning}

Our preprocessing acts on classical data before encoding, aiming to increase the encoded states' mean algebraic purity.
Rotating the data changes their orientation relative to the coordinate planes from which the encoding angles are extracted.
We first consider a nonzero input $\bx\in\mathbb{R}^D$, where $D\ge2n$, and apply a Haar-random rotation $Q\in SO(D)$ that is subsequently held fixed across the dataset.
The $i$-th encoding angle is the polar angle of the $i$-th disjoint coordinate plane, $\phi_i=\operatorname{atan2}\big((Q\bx)_{2i},(Q\bx)_{2i-1}\big)\bmod 2\pi$.
This rotation is motivated by classical input preprocessing~\cite{lecun_efficient_2012}, although an orthogonal transformation alone does not whiten the data.
For each fixed nonzero input, averaging over $Q$ makes these $n$ angles independent and uniform on $[0,2\pi)$ (\autoref{apx:input_transformation}).
This per-input statement uses $D\ge2n$ disjoint planes and need not hold if coordinates are reused.
If necessary, padding a nonzero input before drawing $Q$ in the enlarged space retains the per-input uniform law.

For a dataset, the same rotation $Q$ acts on every input.
One shared rotation need not produce independent uniform angles across observed inputs, so we test the resulting empirical mean purity in~\autoref{apx:input_transformation}.
To do so, we draw a candidate rotation, evaluate this mean from the off-diagonal closed form, and retain the rotation only if it passes the threshold.
Once accepted, this rotation is held fixed, so the encoding remains a deterministic function of the input.
In addition, extracting polar angles discards the radii of the coordinate pairs and, when $D>2n$, any unused coordinates.
Thus, although $Q$ is invertible, the complete map need not preserve information required by the learning task.

\paragraph{Circuits}

Our two baseline circuits have polynomial-dimensional \glspl{dla}, and one layer of each \as is shown in~\autoref{fig:ansaetze}.
The \gls{mga} on an open chain is generated by $\{Z_k\}_{k=1}^{n}\cup\{X_kX_{k+1}\}_{k=1}^{n-1}$ and has $\mg\cong\mso(2n)$ with dimension $n(2n-1)$.
It is classically simulable~\cite{jozsa_matchgates_2008}.
Replacing the single-site generators with nearest-neighbor $Y_kY_{k+1}$ gives the off-diagonal \as, generated by $\{X_kX_{k+1},Y_kY_{k+1}\}$, whose algebra is $\god\cong\mso(n)\oplus\mso(n)$ with dimension $n(n-1)$~\cite{wiersema_classification_2024}.
To examine larger algebras, we also add the generators $\{X_kY_{k+2}\}_{k=1}^{n-2}$ or place the $XY$ pair $\{X_iX_j,Y_iY_j\}$ on the edges of a connected graph with a vertex of degree greater than two.
These graph gate sets support encoded universal computation and have exponential \glspl{dla}~\cite{brod_computational_2014,kokcu_classification_2026}.
Their dimension removes the efficiency guarantee of simulation methods that track the full algebra~\cite{goh_lie_2025}, but it does not establish hardness for the particular product inputs and readouts used here.
For the doped chain, the formula $\dim\gh=4^{n-1}-2^{n-1}$ is verified numerically through $n=9$ and conjectured beyond that range.
A hardness reduction for this gate set remains open.

\paragraph{Encoding and Measurement}

All families use the product encoding $\rho(\bphi)=\ketbra{\psi(\bphi)}{\psi(\bphi)}$, where $\ket{\psi(\bphi)}=\bigotimes_{i=1}^{n}\RY(\phi_i)\ket{0}$.
We use the rotation convention $R_P(\alpha)=\exp(-\imag\alpha P/2)$ for a Pauli string $P$.
We analyze a \gls{qfm}~\cite{schuld_effect_2021} with one encoding block before turning to data re-uploading in the numerical experiments.
Unless a comparison caption specifies otherwise, the matchgate baseline uses $\obs=Z_i$, while the off-diagonal training run and variance comparisons use $\obs=X_iX_{i+1}+Y_iY_{i+1}$.
The doped-chain validation also uses $\obs=X_1X_2$.
These observables lie in their respective algebras, allowing us to evaluate their projections in~\autoref{eq:variance}.
By comparison, single-site $Z_i$ lies outside the floor-free algebras, so that formula does not describe its variance there.

\autoref{tab:notation} collects the recurring symbols.

\begin{table}[!htb]
  \centering
  \footnotesize
  \caption{Recurring notation}
  \label{tab:notation}
  \begin{tblr}{colspec={l X[l]}, colsep=4.0pt, width=\linewidth, row{1}={font=\bfseries}}
    \toprule
    Symbol & Meaning \\
    \midrule
    $\mg$, $\mg_j$ & \gls{dla}, with simple ideals $\mg_j$ ($j<k$) and abelian center $\mg_k$ \\
    $\dim\mg$ & Dimension of the \gls{dla} \\
    $\pur_\mg(\rho)$ & Pauli-basis $\mg$-purity, $=2^n\pur^{\mathrm{HS}}_\mg(\rho)$ \\
    $d_A(\mg)$ & Number of Pauli basis strings of $\imag\mg$ diagonal in the eigenbasis of the clustering observable $A$ \\
    $d_Z$, $d_X$ & Its instances under the $\RY$ encoding, for angles clustered at $\{0,\pi\}$ ($A=Z$) and at $\nicefrac{\pi}{2}$ ($A=X$) \\
    $\god$, $\gh$ & Off-diagonal $\mso(n)\oplus\mso(n)$ and doped-chain algebra \\
    $\bphi$, $\btheta$ & Encoding angles and trainable parameters \\
    $\obs$ & Measurement observable, in-algebra with $\obs\in \imag\mg$ \\
    $n$, $D$, $m$ & Numbers of qubits, classical coordinates, and dataset samples \\
    $Q$, $\mu_n$ & Shared rotation and uniform-prior mean of $\pur_\god$ \\
    $\sigma$, $\bw$ & Angular jitter standard deviation and scalar-encoding weight vector \\
    $\Omega$ & \gls{qfm} frequency spectrum of size $|\Omega|$ \\
    $\mathcal{O}(\cdot),\Omega(\cdot),\Theta(\cdot)$ & Asymptotic bounds in $n$ \\
    \bottomrule
  \end{tblr}
\end{table}

\section{Results}
\label{sec:results}

We first derive purity expressions for the matchgate and off-diagonal families, then examine what changes when the encoding or circuit algebra is enlarged.
Unless stated otherwise, purity curves evaluate the closed forms below, and prior means use $4\times10^3$ angle draws.
The uniform prior draws angles independently from $\mathcal{U}[0,2\pi)$.
The clustered prior uses independent angles $\phi_k=(\pi b_k+\epsilon_k)\bmod 2\pi$, with independent $b_k\sim\operatorname{Bernoulli}(1/2)$ and $\epsilon_k\sim\mathcal{N}(0,\sigma^2)$.
Thus, $\sigma$ measures the standard deviation of the angular jitter in radians.
The re-uploading depth sweep instead clusters all angles around zero, as specified in~\autoref{fig:reupload}.
We check selected formulas by summing over the full Pauli-string basis and compare variance predictions with deep random circuits in statevector simulation.
For these numerical experiments, we use QML-Essentials~\cite{strobl_qml_2025,franz_software_2026}.
The code to reproduce results and figures of this work is available online.\footnote{\geneurl{https://github.com/cirKITers/unflattening-by-flattening}}

\paragraph{Floored Regime}

\begin{figure}[tb]
  \centering
  \input{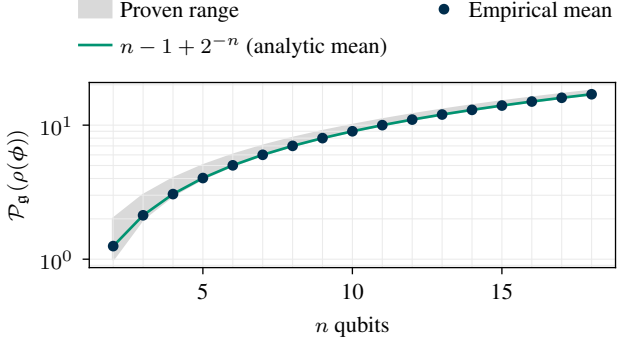}
  \caption{Under the uniform prior, the empirical mean of $\pur_\mg(\rho(\bphi))$ matches $n-1+2^{-n}$ and stays within $[n-1,\,n]$, for sizes up to $n=18$ qubits.}
  \label{fig:bmean}
\end{figure}

The matchgate family provides a baseline in which input purity varies little with encoding angles.
For any pure product state and $\mg\cong\mso(2n)$, this purity has the closed form
\begin{equation}
  \pur_\mg(\rho) = (n-1) + \prod_{k=1}^{n}\braket{Z_k}^2 ,
  \label{eq:productform}
\end{equation}
For the $\RY$ loading, $\braket{Z_k}=\cos\phi_k$.
For a product state, each Pauli-string expectation factorizes across sites.
The matchgate basis consists of the single-site $Z_k$ and the two-endpoint strings $\sigma_j(\prod_{j<l<k}Z_l)\sigma'_k$, with $\sigma,\sigma'\in\{X,Y\}$~\cite[Supplementary Eq.~(97)]{ragone_lie_2024}.
Summing their squared expectations telescopically yields~\autoref{eq:productform}.
Because the product term is the squared global-parity expectation, the purity lies in $[n-1,n]$ for every angle configuration and has uniform-prior mean $n-1+2^{-n}$ near the lower end of this band.
Brute-force evaluations through $n=8$ agree with the closed form to machine precision, and~\autoref{fig:bmean} shows the corresponding uniform-prior mean.

For the readout $Z_i$ and $n\ge3$, \autoref{eq:variance} therefore gives $\var_{\btheta}[f_{\btheta}]=\pur_\mg/\dim\mg=\Theta(\nicefrac{1}{n})$ for every pure product input.
This guarantee holds separately for each input, without averaging over an angle distribution.
Within this circuit ensemble, changing that distribution cannot alter the asymptotic scaling.
Preconditioning is therefore unnecessary for protecting this scaling in the matchgate baseline.
Throughout, $\mathcal{O}(\cdot)$, $\Omega(\cdot)$ and $\Theta(\cdot)$ denote asymptotic upper, lower and two-sided bounds in $n$.

\paragraph{Floor Criterion}

The narrow matchgate band suggests asking what can be learned from the algebra when encoded inputs cluster.
At a computational-basis input, a Pauli string containing an $X$ or $Y$ factor has zero expectation, while a string made only of $I$ and $Z$ has expectation $\pm1$.
Counting the surviving strings therefore reveals which part of the input purity remains at this endpoint.
Suppose each encoding factor approaches an eigenstate of a single-qubit Pauli observable $A$, and let $d_A(\mg)$ count the Pauli basis strings of $\imag\mg$ that are diagonal in the $A$-eigenbasis.
This count depends on both the clustering basis and the algebra.
For $\RY$ loading, clustering at $\{0,\pi\}$ gives $d_Z$, whereas clustering at $\nicefrac{\pi}{2}$ gives $d_X$.
Indeed, the off-diagonal algebra has $d_Z(\god)=0$ but $d_X(\god)=n-1$.

At a product input $\ket{a}$ of $A$-eigenstates, the purity equals this count exactly, $\pur_\mg(\ketbra{a}{a})=d_A(\mg)$, regardless of which eigenstates are chosen.
For any Pauli basis string $B=\bigotimes_kB_k$, its expectation factorizes as $\braket{B}_{\ket{a}}=\prod_k\braket{B_k}_{a_k}$.
Each factor belonging to $\{I,A\}$ has magnitude one, while a factor anticommuting with $A$ has zero expectation.
Consequently, only the $d_A(\mg)$ strings built from $I$ and $A$ survive after squaring and summing their expectations.
We call the resulting clustered value the \emph{floor} of the encoding-algebra pair.
The term refers to the clustered endpoint, rather than a lower bound for every input.
For matchgates, the uniform-prior mean $n-1+2^{-n}$ lies below $d_Z=n$.
For matchgates, the full band in~\autoref{eq:productform} shows why purity varies little with the input distribution.

We call an encoding-algebra pair \emph{floor-free} when $d_A(\mg)=0$.
Its in-algebra output is zero at the exact clustered endpoint for every parameter realization, regardless of depth or $\dim\mg$ (\autoref{eq:exact-zero}).
At nonzero clustering width, the degree of suppression instead depends on both the width and system size.
For $\RY$ encoding, $A=Z$ at the computational-basis endpoint, while the single-site $Z_k$ strings give the matchgate family $d_Z=n$.

This count alone does not determine output variance.
In our Pauli-basis convention, each surviving string contributes one, whereas its contribution in the \gls{hs} normalization is $2^{-n}$.
Moreover, \autoref{eq:variance} weights the input projection onto each simple ideal by the corresponding observable projection.
A simple ideal need not itself have a Pauli-string basis, even when the full algebra does, so its contribution cannot generally be replaced by an integer string count.
Diagonal strings in the abelian center also enter $d_A$ without contributing to variance.
For example, adjoining the central string $Z^{\otimes n}$ to $\god$ gives $d_Z=1$ but leaves the output parameter-independent on computational-basis inputs.
If the readout has a central component, that constant can be nonzero.

\autoref{fig:criterion} contrasts the matchgate purity band with the collapse of the off-diagonal purity as angles approach the computational basis.

\begin{figure}[tb]
  \centering
  \input{figures/purity_regime_contrast.pgf}
  \caption{Purity and output-variance response to angle clustering at $n=10$ under the $\RY$ encoding.
    As the angle spread $\sigma$ shrinks toward the computational basis, the mean matchgate purity stays in $[n-1,n]$, whereas the mean off-diagonal purity approaches zero.
    The right axis (dashed) traces the mean sample-wise output variance $\expphi[\var_{\btheta}[f_{\btheta}]]=2\expphi[\pur_\mg]/\dim\mg$ for the in-algebra observable $X_iX_{i+1}+Y_iY_{i+1}$ under the ensemble of~\autoref{eq:variance}.
    Accordingly, the matchgate variance changes little, while the off-diagonal variance approaches zero with its purity.}
  \label{fig:criterion}
\end{figure}
\paragraph{Floor-Free Polynomial Regime}

By contrast, the off-diagonal \as has polynomial dimension but no $Z$-diagonal string, so it satisfies $d_Z(\god)=0$ under $\RY$ encoding.
Its Pauli basis contains $XX$ and $YY$ endpoints at odd separation and $XY$ and $YX$ endpoints at even separation, joined by interior $Z$ factors~\cite{wiersema_classification_2024}.
Since $\braket{Y_k}=0$ for the $\RY$ product encoding, only the $XX$ strings contribute, giving
\begin{equation}
  \pur_\god\!\left(\rho(\bphi)\right)
  = \sum_{\substack{j<k \\ k-j\ \mathrm{odd}}} \sin^2\phi_j\,\sin^2\phi_k \!\!\prod_{l=j+1}^{k-1}\!\! \cos^2\phi_l .
  \label{eq:offdiag-closedform}
\end{equation}

Every term contains $\sin^2\phi_j\sin^2\phi_k$, so the purity approaches zero as all angles cluster near $\{0,\pi\}$.
At the other clustering point, $\phi_k=\nicefrac{\pi}{2}$ for every site, the inputs become $X$-eigenstates and attain the global maximum $\pur_\god=n-1$ (\cf~\autoref{apx:purity}).
This endpoint agrees with $d_X(\god)=n-1$: only the adjacent $X_kX_{k+1}$ basis strings are diagonal in the $X$ basis, since longer strings contain interior $Z$ factors.
Between these endpoints, the \iid uniform prior gives
\begin{equation}
  \expphi\!\left[ \pur_\god\!\left(\rho(\bphi)\right) \right]
  = \sum_{\substack{1\le d<n \\ d\ \mathrm{odd}}} (n-d)\,2^{-(d+1)}
  = \Theta(n),
  \label{eq:offdiag-mean}
\end{equation}

We denote this mean by $\mu_n$, with $\mu_n=n/3-5/9+\mathcal{O}(2^{-n})$.
Thus, uniform angles give mean purity of order $n$, within a constant factor of the maximum.
This mean does not guarantee large purity for each input or with high probability.
For $n=3$ or $n\ge5$, the readout $X_iX_{i+1}+Y_iY_{i+1}$ places one basis string in each of the two equal-dimensional simple ideals of $\god$.
\autoref{eq:variance} then gives $\var_{\btheta}[f_{\btheta}]=2\pur_\god/\dim\god$ (\cf~\autoref{apx:purity}), so the uniform prior yields $\expphi[\var_{\btheta}[f_{\btheta}]]=\Theta(\nicefrac{1}{n})$.
At $n=4$, the same aggregate formula holds, with the four simple ideals accounted for in~\autoref{apx:purity}.
Thus, the same circuit and readout have exactly zero output at computational-basis inputs and inverse-polynomial mean output variance under uniform angles.

For a fixed dataset, the uniform-prior calculation also supplies a concrete preprocessing target.
Under the conditions of~\autoref{sec:background}, the acceptance test in~\autoref{apx:input_transformation} retains a shared rotation when the empirical mean purity reaches at least half of the uniform-prior mean.
Every accepted rotation certifies empirical mean sample-wise output variance $\Theta(\nicefrac{1}{n})$ for the specified readout and circuit ensemble.
Moreover, rejection sampling needs only a constant expected number of candidate rotations, with the bound approaching five as $n$ grows.
This guarantee applies to any fixed dataset satisfying the assumptions, without requiring independent or uniformly distributed examples.

\begin{figure}[tb]
  \centering
  \input{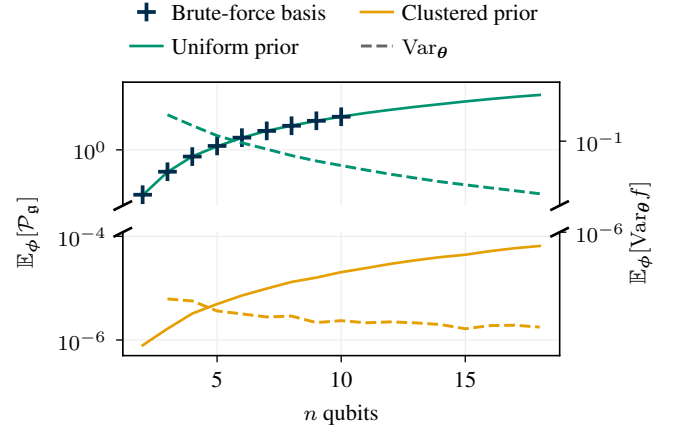}
  \caption{Mean purity and sample-wise output variance for the off-diagonal algebra $\god\cong\mso(n)\oplus\mso(n)$, which has $d_Z=0$.
    On the left axis, the closed form~\autoref{eq:offdiag-closedform} agrees with brute-force values (markers) and gives linear growth under the uniform prior, whereas clustering near $\{0,\pi\}$ suppresses the mean by about five orders of magnitude.
    The dashed curves on the right axis show the corresponding mean variance $2\expphi[\pur_\god]/\dim\god$ for $n\ge3$.
    At $n=2$ the algebra is abelian, so its zero output variance is omitted from the logarithmic axis.}
  \label{fig:offdiag}
\end{figure}

Uniform angles are sufficient for the stated mean scaling but need not be optimal for a learning task.
For example, mapping every input to the purity-maximizing configuration would erase distinctions among examples, while polar-angle extraction can also discard task-relevant information.
\autoref{fig:offdiag} compares the closed form with brute-force checks for $n\le6$ and shows the two prior means.
With fixed jitter $\sigma=0.03$, the uniform-to-clustered mean purity ratio decreases from about $2.0\times10^5$ at $n=6$ to $8.3\times10^4$ at $n=18$.
Although suppression is substantial over this range, these values do not establish exponential decay with $n$.
At $n=6$, a statevector check also agrees with the plotted variance, including its factor of two, within sampling error.

\paragraph{Data Re-Uploading}

\begin{figure}[tb]
  \centering
  \input{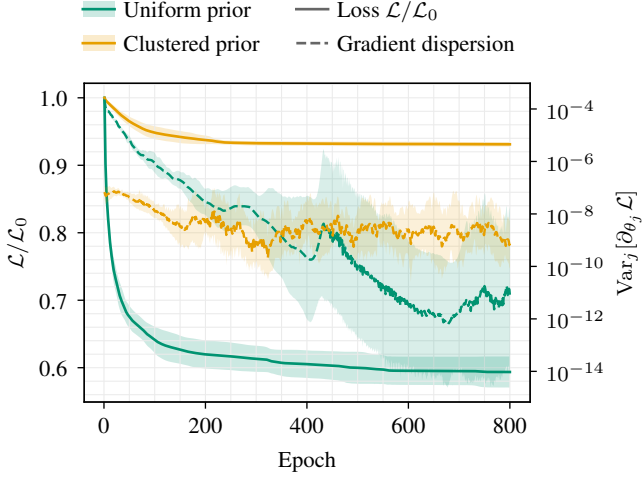}
  \caption{Training the off-diagonal re-uploading model under uniform and clustered ($\sigma=0.03$) angle priors.
    Each run's relative loss is $\loss(\btheta_t)/\loss(\btheta_0)$.
    Solid lines show its mean across $8$ paired initializations, with bands of $\pm1$ standard deviation.
    Dashed lines show gradient dispersion across parameter coordinates, $\var_j[\partial_{\theta_j}\loss]$, at each training step.
    Their central curves and bands are computed as $\exp(\operatorname{mean}(\log v))$ and $\exp(\operatorname{mean}(\log v)\pm\operatorname{std}(\log v))$ across seeds, where $v$ is the dispersion.
    This dispersion is calculated across coordinates rather than across random parameter initializations at a fixed circuit ensemble.}
  \label{fig:precondition_training}
\end{figure}

To see whether this contrast appears during training, we compare optimization with Adam under two input priors in the data re-uploading model of~\autoref{fig:precondition_training}.
The $n=6$ circuit uses the readout $\obs=X_iX_{i+1}+Y_iY_{i+1}$ and begins with a trainable block.
It contains $10$ $\RY$ encoding blocks alternating with $11$ off-diagonal trainable blocks.
We draw one fixed dataset of $256$ inputs from each prior and label both with the same Fourier-series target function.
This target uses frequencies in $\{-1,0,1\}^n$ with at most two nonzero entries and is standardized using $4\times10^3$ independent uniform reference inputs.
For each of $8$ seeds, the two runs start from the same parameters, sampled independently from $\mathcal{U}[0,2\pi)$, and minimize $\loss(\btheta)=\frac{1}{256}\sum_{a=1}^{256}(f_{\btheta}(\bphi^{(a)})-y_a)^2$ at learning rate $0.05$ for $800$ epochs.
Here $f_{\btheta}(\bphi^{(a)})$ denotes the complete re-uploading circuit's output on sample $a$.
Because the labeled datasets differ and neither is obtained by rotating the other, this experiment compares input priors rather than preprocessing the same dataset.

The uniform-input runs finish at relative loss $0.594\pm0.023$, compared with $0.931\pm0.003$ for the clustered-input runs.
These correspond to decreases of about $41\%$ and $7\%$, respectively, in this finite experiment.
However, the plotted gradient dispersion changes order during training and does not establish \gls{bp} scaling.

The purity formulas do not directly describe this training circuit.
\autoref{eq:productform} and~\autoref{eq:offdiag-closedform} apply to an encoded product state before a trainable block, whereas this circuit starts with a trainable block.
Its product-state purity is therefore a reference diagnostic, not the actual purity at the first trainable block.
Later encoding layers act on states that generally depend on $\btheta$ and may be entangled.
Because the $\RY$ generators lie outside $\imag\god$, \autoref{eq:variance} does not directly describe the full multi-block circuit.
The exact computational-basis case remains covered by~\autoref{eq:exact-zero}, while the behavior at nonzero width and under uniform angles is supported by finite-depth observations such as~\autoref{fig:reupload}.

\paragraph{Weighted Scalar Encodings}

When all angles come from one scalar input, the encoding itself can also shape their distribution.
For $\phi_k=w_kx$ with integer weights $\bw=(w_1,\ldots,w_n)$, a single encoding block has the allowed \gls{qfm} frequency set $\Omega=\{\sum_k s_kw_k\mid s_k\in\{-1,0,1\}\}$~\cite{schuld_effect_2021}.
However, individual coefficients may vanish, and repeated encoding blocks generally enlarge this set of allowed frequencies.

Equal (Hamming) weights $w_k=1$ give $|\Omega|=2n+1$, whereas binary weights $w_k=2^{k-1}$ and ternary weights $w_k=3^{k-1}$ give $|\Omega|=2^{n+1}-1$ and $3^n$, respectively~\cite{peters_generalization_2023,shin_exponential_2023}.
Because all angles share one scalar input, $\pur_\mg(\rho(\bw x))$ becomes a trigonometric polynomial with frequencies in $2\Omega$, producing the oscillatory landscapes illustrated in~\autoref{fig:encodings-landscape}.

To characterize their mean purity, we call the weights \emph{dissociated} when $\sum_k\epsilon_kw_k\neq0$ for every nonzero $\bepsilon\in\{-1,0,1\}^n$~\cite{tao_additive_2006}.
Binary and ternary weights satisfy this condition.
For the $\RY$ product encoding, squared Pauli expectations are products of $\sin^2\phi_k$ and $\cos^2\phi_k$, whose Fourier expansions contain frequencies $2\sum_k\epsilon_kw_k$~\cite{schuld_effect_2021}.
Dissociation rules out a zero frequency for every nonconstant term, so averaging over uniform $x$ retains the same constant term as averaging over independent uniform angles~\cite{tao_additive_2006}:
\begin{equation}
  \mathbb{E}_x\!\left[\pur_\mg\!\left(\rho(\bw x)\right)\right]
  \;=\; \expphi\!\left[\pur_\mg\!\left(\rho(\bphi)\right)\right],
  \label{eq:spectral-precondition}
\end{equation}
for dissociated $\bw$ and $x\sim\mathcal{U}[0,2\pi)$.
Thus, one uniformly distributed scalar can reproduce the mean input purity obtained from $n$ independent uniform angles.
We refer to this equality of means as \emph{spectral preconditioning}.
It does not imply that the induced angles are independent or that purity concentrates near the mean, and it need not hold for equal weights or nonuniform scalar inputs.

This weight condition is related to, but weaker than, conditions used to obtain distinct Fourier frequencies in \glspl{qfm}~\cite{peters_generalization_2023,shin_exponential_2023}.
Binary weights are superincreasing in the usual sense, $w_k>\sum_{j<k}w_j$, yet fail the stronger sufficient condition $w_k>2\sum_{j<k}w_j$ for uniqueness of signed-sum frequencies.
For example, $w_2-w_1=w_1$, so distinct sign patterns can share a frequency even though no nonzero signed subset sums to zero.
Ternary weights, by contrast, satisfy the stronger condition.
The mean-purity identity~\autoref{eq:spectral-precondition} complements the gradient-variance bounds of~\citeauthor{barthe_gradients_2024}~\cite{barthe_gradients_2024}, which do not impose this weight condition.
Large spectra can also have exponentially concentrated Fourier coefficients, so this mean-purity identity alone does not guarantee trainability~\cite{mhiri_constrained_2025}.

\autoref{fig:encodings-mean} compares these encodings through the mean off-diagonal purity $\mathbb{E}_x[\pur_\god]$ under uniform and clustered scalar inputs.
The clustered scalar prior is $x=(\pi b+\epsilon)\bmod 2\pi$, with independent $b\sim\operatorname{Bernoulli}(1/2)$ and $\epsilon\sim\mathcal{N}(0,\sigma^2)$.

\begin{figure}[tb]
  \centering
  \input{figures/encoding_mean.pgf}
  \caption{Mean purity $\mathbb{E}_x[\pur_\god]$ against $n$ under uniform (solid) and clustered (dashed, $\sigma=0.03$) scalar inputs, estimated from $2\times10^5$ draws of $x$ ($\god$, $\RY$ encoding).
    Empirical means for dissociated binary and ternary weights agree with the \iid mean under uniform $x$ and partially recover under clustered $x$, whereas Hamming weights remain strongly suppressed.
    The dotted right-axis curves show the allowed frequency-set sizes $|\Omega|$: $2n+1$ for equal weights, $2^{n+1}-1$ for binary weights, and $3^n$ for ternary weights.}
  \label{fig:encodings-mean}
\end{figure}

Under uniform $x$, the numerical means agree with~\autoref{eq:spectral-precondition} within sampling error.
Under the clustered scalar prior, larger weights amplify the angular jitter at individual sites.
In the measured range, this brings the binary-weight mean within a factor of two of its uniform value, while the equal-weight mean remains several orders of magnitude smaller.
This recovery at finite jitter is observed only for the tested priors.

\paragraph{Extensions to Exponential Algebras}

The polynomial-dimensional examples show how input alignment affects purity, but the same clustered endpoint can occur in larger algebras.
The bipartite $XY$ graph families and the $X_kY_{k+2}$-doped chain in~\autoref{fig:hard-graphs} have $d_Z=0$ through the tested sizes $n=8$ and $n=9$, respectively.
For the graph families, the exponential algebra dimensions follow Ref.~\cite{kokcu_classification_2026}.
For the doped chain, $\dim\gh=4^{n-1}-2^{n-1}$ is verified numerically over the tested range, where the dopant is a reusable gate generator rather than a single gate insertion.
Adding an odd chord to the graph changes the result: the observed diagonal count becomes $d_Z=2^{n-1}-2$ at even $n$ and $2^{n-1}-1$ at odd $n$, for $n\le8$.

When $d_Z=0$ and $\obs\in \imag\mg$, the consequence at a computational-basis input $\ket{a}$ is especially direct.
For every circuit $U$ generated by $\mg$, regardless of depth or parameter values,
\begin{equation}
  \braket{a|U^\dagger \obs U|a} \;=\; 0 .
  \label{eq:exact-zero}
\end{equation}

\autoref{eq:exact-zero} holds for each circuit and requires no design assumption.
At this endpoint, increasing the trainable depth cannot restore a parameter-dependent output.
If the training objective is mean squared error, the resulting constant loss equals the mean squared target value and need not vanish.
This per-circuit statement complements the ensemble variance formula~\cite{ragone_lie_2024,fontana_characterizing_2024} and uses the same adjoint invariance relevant to Lie-algebraic simulation~\cite{goh_lie_2025}.

The endpoint criterion identifies inputs with exactly zero output.
For other inputs, the ensemble variance measures how much the output changes across sampled circuits.
For these other inputs, the attainable variance also depends on the algebra's component dimensions.
The bound $\pur_\mg(\rho)\le2^n$ gives variance $\mathcal{O}(2^{-n})$ through~\autoref{eq:variance} when every observable-relevant simple ideal has dimension $\Theta(4^n)$ and the observable has bounded operator norm.
These branched graph families have the required component scaling~\cite{kokcu_classification_2026}, whereas the corresponding structure of the doped chain remains a numerical inference.
This condition concerns the relevant simple ideals, not merely the dimension of the full algebra.
It also presumes Haar or exact 2-design sampling, or an approximation accurate enough to preserve the scaling.
Structured initializations, however, need not satisfy this ensemble assumption.
An exponential algebra alone does not determine the required design depth, since polynomial-depth design constructions are known~\cite{brandao_local_2016}.
We therefore leave the mixing depth of the gate schedules studied here unresolved.

\begin{figure}[tb]
  \centering
  \input{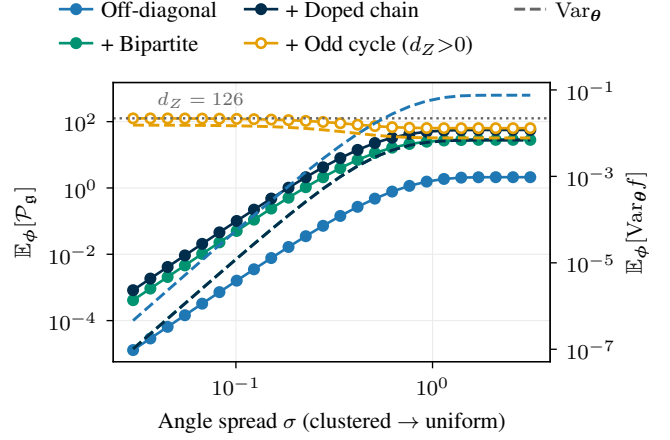}
  \caption{Purity and predicted Haar output variance for the off-diagonal chain $\god$ and three perturbations at $n=8$.
    Their interaction graphs appear in~\autoref{fig:hard-graphs}.
    Only the odd cycle has $d_Z>0$, while the other three families have $d_Z=0$.
    Solid curves on the left axis give the exact mean purity for angles clustered near $\{0,\pi\}$ with spread $\sigma$.
    The floored odd cycle stays pinned near its clustered limit $d_Z=2^{n-1}-2=126$ (dotted), whereas the floor-free purities collapse and reach exactly zero at $\sigma=0$ (\autoref{eq:exact-zero}).
    Dashed curves on the right axis give the mean Haar output variance for $X_1X_2+Y_1Y_2$.
    At this size, the observable has equal \gls{hs} weight in the equal-dimensional simple ideals of each family, giving $2\expphi[\pur_\mg]/\dim\mg$.
    There are two ideals for the chain and, numerically, the doped chain.
    Each graph family has four ideals at this size~\cite{kokcu_classification_2026}.
    The larger component dimensions reduce this variance relative to the chain.}
  \label{fig:hollow}
\end{figure}

\autoref{fig:hollow} compares the finite-size purity responses with the corresponding Haar-variance predictions.
At computational-basis angles, each encoding layer is a Pauli string up to a global phase.
Conjugation by a Pauli string only changes the signs of Pauli basis operators, so it preserves $\imag\mg$~\cite[Appendix~A]{kokcu_classification_2026}.
Together with adjoint invariance under the trainable blocks~\cite{ragone_lie_2024}, this preserves the zero-output identity at any re-uploading depth when the circuit starts from a computational-basis state.
For uniform inputs, the tractable-chain measurements in~\autoref{fig:reupload} provide a finite-depth comparison, while general re-uploading gradients require separate arguments~\cite{barthe_gradients_2024}.

The floor-free perturbations remain sensitive to clustering over the tested sizes, while the odd-cycle control retains nonzero purity at the exact endpoint.
For the doped chain at $n=6$, deep-circuit checks with readout $X_1X_2$ give empirical-to-Haar variance ratios of $1.06$ and $1.05$ for two uniform input draws and $0.91$ for a draw clustered around zero with $\sigma=0.15$.
These finite checks support the plotted predictions without measuring an ensemble-wide design error.

\section{Conclusion}
\label{sec:outlook}

Input distributions control how encoded states overlap with the operators available to the circuit and readout.
For the matchgate family, every pure product input retains enough overlap to give output variance $\Theta(\nicefrac{1}{n})$ with the specified readout under Haar or exact group 2-design sampling.
For the off-diagonal family, computational-basis inputs instead give exactly zero output for every parameter choice, while independent uniform angles yield mean sample-wise variance $\Theta(\nicefrac{1}{n})$ under the same ensemble assumption.
Counting diagonal strings explains these different clustered endpoints and provides a direct way to identify inputs with zero in-algebra output.
\autoref{tab:summary} collects the resulting variance scalings and their assumptions.

\begin{table*}[!htb]
  \centering
  \footnotesize
  \caption{Summary for the $\RY$ product encoding under the Haar or exact group 2-design ensemble.
    The uniform column reports mean sample-wise output variance, and the final column reports the exact computational-basis endpoint $\bphi\in\{0,\pi\}^n$.
    It does not describe a prior with finite clustering width.
    Daggers mark diagonal counts checked through $n=8$ for graphs and $n=9$ for the doped chain.
    The doped dimension and its asymptotic variance cap also remain conditional on the conjectured component structure.}
  \label{tab:summary}
  \begin{tblr}{colspec={lccccc}, colsep=6.0pt, row{1}={font=\bfseries\boldmath}}
    \toprule
    Family & $\dim\mg$ & $d_Z$ & $\obs$ & $\expphi[\var_{\btheta}[f_{\btheta}]]$ (Uniform) & $\var_{\btheta}[f_{\btheta}]$ (Exact cluster) \\
    \midrule
    Matchgate chain & $n(2n{-}1)$ & $n$ & $Z_i$ & $\Theta(\nicefrac{1}{n})$ & $\Theta(\nicefrac{1}{n})$ \\
    Off-diagonal chain & $n(n{-}1)$ & $0$ & $X_iX_{i+1}{+}Y_iY_{i+1}$ & $\Theta(\nicefrac{1}{n})$ & $0$ \\
    Branched bipartite $XY$ & $\Theta(4^n)$ & $0^\dagger$ & $X_1X_2{+}Y_1Y_2$ & $\mathcal{O}(2^{-n})$ & $0^\dagger$ \\
    Doped chain & $\Theta(4^n)^\dagger$ & $0^\dagger$ & $X_1X_2{+}Y_1Y_2$ & $\mathcal{O}(2^{-n})^\dagger$ & $0^\dagger$ \\
    Odd-cycle $XY$ & $\Theta(4^n)$ & $\Theta(2^n)^\dagger$ & $X_1X_2{+}Y_1Y_2$ & $\mathcal{O}(2^{-n})$ & $\mathcal{O}(2^{-n})$ \\
    \bottomrule
  \end{tblr}
\end{table*}
For the off-diagonal chain, the closed form also turns the uniform-prior calculation into a guarantee for a fixed dataset.
Specifically, the shared-rotation procedure accepts an encoding only after its empirical mean purity passes a computable threshold.
For any dataset satisfying the preprocessing assumptions, a constant expected number of trials suffices to obtain the stated inverse-polynomial mean sample-wise output variance.
The certificate can therefore be checked entirely classically before running the quantum circuit.

Two further results concern the encoding.
The exact zero-output identity persists through computational-basis re-uploading at any depth.
Dissociated binary and ternary weights recover the independent-angle mean purity exactly for uniform scalar inputs.
For the exponential graph families, however, input choice cannot remove the $\mathcal{O}(2^{-n})$ variance cap under the stated ensemble and readout assumptions.
For the doped chain, the analogous asymptotic statement remains conditional on its component structure.
The polynomial families that make these effects analytically accessible are classically simulable~\cite{jozsa_matchgates_2008,goh_lie_2025}.
Their role here is to make this dependence on input distributions explicit and computationally verifiable.

The numerical experiments show substantial suppression at finite clustering width and different optimization trajectories under the two input priors.
These observations do not establish an asymptotic \gls{bp} at fixed width or a general guarantee for supervised learning.
Successful learning also depends on the information retained by the encoding, the labels, how sample outputs vary together, and optimization dynamics.

The diagonal count itself applies to product states clustered in local Pauli eigenbases, with the local basis allowed to vary by site.
Non-product encodings, however, require another analysis.
For mixed states diagonal in the clustering product basis, the count instead gives the upper bound $\pur_\mg\le d_A(\mg)$ because the diagonal expectations need not have unit magnitude.
This bound does not extend to arbitrary noisy states.

Under these assumptions, flattening the angle distribution can raise the off-diagonal model's average output variance, which can be certified for the encoded dataset.
This gives a concrete link between a classical preprocessing choice and the output landscape of a fixed quantum circuit family.

\small
\noindent\textbf{Acknowledgements}
Implementation of numerical experiments, as well as grammar and wording checks were partially assisted by Anthropic Claude Opus series.
All authors acknowledge support by the state of Baden-W\"urttemberg through bwHPC.
\normalsize

\printbibliography

@article{schuld_effect_2021,
  title = {Effect of Data Encoding on the Expressive Power of Variational Quantum-Machine-Learning Models},
  author = {Schuld, Maria and Sweke, Ryan and Meyer, Johannes Jakob},
  year = 2021,
  month = mar,
  journal = {Physical Review A},
  volume = {103},
  number = {3},
  eprint = {2008.08605},
  pages = {032430},
  issn = {2469-9926, 2469-9934},
  doi = {10.1103/PhysRevA.103.032430},
  archiveprefix = {arXiv}
}

@article{ragone_lie_2024,
  title = {A {{Lie Algebraic Theory}} of {{Barren Plateaus}} for {{Deep Parameterized Quantum Circuits}}},
  author = {Ragone, Michael and Bakalov, Bojko N. and Sauvage, Fr{\'e}d{\'e}ric and Kemper, Alexander F. and {Ortiz Marrero}, Carlos and Larocca, Mart{\'i}n and Cerezo, M.},
  year = 2024,
  month = aug,
  journal = {Nature Communications},
  volume = {15},
  number = {1},
  eprint = {2309.09342},
  primaryclass = {quant-ph},
  pages = {7172},
  issn = {2041-1723},
  doi = {10.1038/s41467-024-49909-3},
  urldate = {2025-03-01},
  archiveprefix = {arXiv}
}

@article{fontana_characterizing_2024,
  title = {Characterizing barren plateaus in quantum ansätze with the adjoint representation},
  author = {Fontana, Enrico and Herman, Dylan and Chakrabarti, Shouvanik and Kumar, Niraj and Yalovetzky, Romina and Heredge, Jamie and Sureshbabu, Shree Hari and Pistoia, Marco},
  year = 2024,
  month = aug,
  journal = {Nature Communications},
  volume = {15},
  number = {1},
  pages = {7171},
  issn = {2041-1723},
  doi = {10.1038/s41467-024-49910-w},
  eprint = {2309.07902},
  archiveprefix = {arXiv},
  primaryclass = {quant-ph}
}

@article{cerezo_does_2025,
  title = {Does Provable Absence of Barren Plateaus Imply Classical Simulability?},
  author = {Cerezo, M. and Larocca, Martin and {Garc{\'i}a-Mart{\'i}n}, Diego and Diaz, N. L. and Braccia, Paolo and Fontana, Enrico and Rudolph, Manuel S. and Bermejo, Pablo and Ijaz, Aroosa and Thanasilp, Supanut and Anschuetz, Eric R. and Holmes, Zo{\"e}},
  year = 2025,
  journal = {Nature Communications},
  volume = {16},
  pages = {7907},
  doi = {10.1038/s41467-025-63099-6},
  eprint = {2312.09121},
  archiveprefix = {arXiv},
  primaryclass = {quant-ph}
}

@misc{kulshrestha_exponentially_2026,
  title = {Exponentially Many Initializations to Avoid Barren Plateaus},
  author = {Kulshrestha, Ankit and Puig, Ricard and {Garc{\'i}a-Mart{\'i}n}, Diego and Cincio, Lukasz and Safro, Ilya and Holmes, Zo{\"e} and Cerezo, M.},
  year = 2026,
  month = jun,
  number = {arXiv:2606.18515},
  eprint = {2606.18515},
  archiveprefix = {arXiv},
  primaryclass = {quant-ph},
  publisher = {arXiv},
  doi = {10.48550/arXiv.2606.18515}
}

@article{jozsa_matchgates_2008,
  title = {Matchgates and Classical Simulation of Quantum Circuits},
  author = {Jozsa, Richard and Miyake, Akimasa},
  year = 2008,
  journal = {Proceedings of the Royal Society A},
  volume = {464},
  number = {2100},
  pages = {3089--3106},
  doi = {10.1098/rspa.2008.0189},
  eprint = {0804.4050},
  archiveprefix = {arXiv},
  primaryclass = {quant-ph}
}

@article{larocca_diagnosing_2022,
  title = {Diagnosing Barren Plateaus with Tools from Quantum Optimal Control},
  author = {Larocca, Martin and Czarnik, Piotr and Sharma, Kunal and Muraleedharan, Gopikrishnan and Coles, Patrick J. and Cerezo, M.},
  year = 2022,
  journal = {Quantum},
  volume = {6},
  pages = {824},
  doi = {10.22331/q-2022-09-29-824},
  eprint = {2105.14377},
  archiveprefix = {arXiv},
  primaryclass = {quant-ph}
}

@article{mhiri_constrained_2025,
  title = {Constrained and Vanishing Expressivity of Quantum {{Fourier}} Models},
  author = {Mhiri, Hela and Monbroussou, L{\'e}o and {Herrero-Gonz{\'a}lez}, Mario and Thabet, Slimane and Kashefi, Elham and Landman, Jonas},
  year = 2025,
  journal = {Quantum},
  volume = {9},
  pages = {1847},
  doi = {10.22331/q-2025-09-03-1847},
  eprint = {2403.09417},
  archiveprefix = {arXiv},
  primaryclass = {quant-ph}
}

@article{barnum_generalizations_2003,
  title = {Generalizations of Entanglement Based on Coherent States and Convex Sets},
  author = {Barnum, Howard and Knill, Emanuel and Ortiz, Gerardo and Viola, Lorenza},
  year = 2003,
  journal = {Physical Review A},
  volume = {68},
  number = {3},
  pages = {032308},
  doi = {10.1103/PhysRevA.68.032308},
  eprint = {quant-ph/0207149},
  archiveprefix = {arXiv},
  primaryclass = {quant-ph}
}

@misc{somma_quantum_2005,
  title = {Quantum Computation, Complexity, and Many-Body Physics},
  author = {Somma, Rolando D.},
  year = 2005,
  note = {PhD thesis},
  eprint = {quant-ph/0512209},
  doi = {10.48550/arXiv.quant-ph/0512209},
  archiveprefix = {arXiv},
  primaryclass = {quant-ph}
}

@article{wiersema_classification_2024,
  title = {Classification of Dynamical {Lie} Algebras of 2-Local Spin Systems on Linear, Circular and Fully Connected Topologies},
  author = {Wiersema, Roeland and K{\"o}kc{\"u}, Efekan and Kemper, Alexander F. and Bakalov, Bojko N.},
  journal = {npj Quantum Information},
  volume = 10,
  number = 1,
  pages = 110,
  year = 2024,
  doi = {10.1038/s41534-024-00900-2},
  eprint = {2309.05690},
  archiveprefix = {arXiv},
  primaryclass = {quant-ph}
}

@article{thanasilp_subtleties_2023,
  title = {Subtleties in the Trainability of Quantum Machine Learning Models},
  author = {Thanasilp, Supanut and Wang, Samson and Nghiem, Nhat A. and Coles, Patrick J. and Cerezo, M.},
  year = 2023,
  journal = {Quantum Machine Intelligence},
  volume = {5},
  number = {1},
  pages = {21},
  doi = {10.1007/s42484-023-00103-6},
  eprint = {2110.14753},
  archiveprefix = {arXiv},
  primaryclass = {quant-ph}
}

@article{mcclean_barren_2018,
  title = {Barren Plateaus in Quantum Neural Network Training Landscapes},
  author = {McClean, Jarrod R. and Boixo, Sergio and Smelyanskiy, Vadim N. and Babbush, Ryan and Neven, Hartmut},
  year = 2018,
  journal = {Nature Communications},
  volume = {9},
  number = {1},
  pages = {4812},
  doi = {10.1038/s41467-018-07090-4},
  eprint = {1803.11173},
  archiveprefix = {arXiv},
  primaryclass = {quant-ph}
}

@article{cerezo_cost_2021,
  title = {Cost Function Dependent Barren Plateaus in Shallow Parametrized Quantum Circuits},
  author = {Cerezo, M. and Sone, Akira and Volkoff, Tyler and Cincio, Lukasz and Coles, Patrick J.},
  year = 2021,
  journal = {Nature Communications},
  volume = {12},
  number = {1},
  pages = {1791},
  doi = {10.1038/s41467-021-21728-w},
  eprint = {2001.00550},
  archiveprefix = {arXiv},
  primaryclass = {quant-ph}
}

@incollection{lecun_efficient_2012,
  title = {Efficient {{BackProp}}},
  author = {LeCun, Yann A. and Bottou, L{\'e}on and Orr, Genevieve B. and M{\"u}ller, Klaus-Robert},
  year = 2012,
  booktitle = {Neural Networks: Tricks of the Trade},
  series = {Lecture Notes in Computer Science},
  volume = {7700},
  pages = {9--48},
  publisher = {Springer},
  doi = {10.1007/978-3-642-35289-8_3}
}

@article{brod_computational_2014,
  title = {The Computational Power of Matchgates and the {{XY}} Interaction on Arbitrary Graphs},
  author = {Brod, Daniel J. and Childs, Andrew M.},
  year = 2014,
  journal = {Quantum Information and Computation},
  volume = {14},
  number = {11--12},
  pages = {901--916},
  doi = {10.26421/QIC14.11-12-1},
  eprint = {1308.1463},
  archiveprefix = {arXiv},
  primaryclass = {quant-ph}
}

@article{kokcu_classification_2026,
  title = {Classification of Dynamical {Lie} Algebras Generated by Spin Interactions on Undirected Graphs},
  author = {K{\"o}kc{\"u}, Efekan and Wiersema, Roeland and Kemper, Alexander F. and Bakalov, Bojko N.},
  year = 2026,
  journal = {Journal of Mathematical Physics},
  volume = {67},
  number = {5},
  pages = {052205},
  eprint = {2409.19797},
  archiveprefix = {arXiv},
  primaryclass = {quant-ph},
  doi = {10.1063/5.0283277}
}

@article{goh_lie_2025,
  title = {Lie-Algebraic Classical Simulations for Quantum Computing},
  author = {Goh, Matthew L. and Larocca, Martin and Cincio, Lukasz and Cerezo, M. and Sauvage, Fr{\'e}d{\'e}ric},
  year = 2025,
  journal = {Physical Review Research},
  volume = {7},
  pages = {033266},
  doi = {10.1103/3y65-f5w6},
  eprint = {2308.01432},
  archiveprefix = {arXiv},
  primaryclass = {quant-ph}
}

@article{shin_exponential_2023,
  title = {Exponential Data Encoding for Quantum Supervised Learning},
  author = {Shin, S. and Teo, Y. S. and Jeong, H.},
  year = 2023,
  month = jan,
  journal = {Physical Review A},
  volume = {107},
  number = {1},
  pages = {012422},
  doi = {10.1103/PhysRevA.107.012422},
  eprint = {2206.12105},
  archiveprefix = {arXiv},
  primaryclass = {quant-ph}
}

@InProceedings{franz_software_2026,
  author = { Franz, Maja and Strobl, Melvin and Hunz, Jonathan and Scheller, Lukas and van der Horst, Lucas and Kuehn, Eileen and Streit, Achim and Mauerer, Wolfgang },
  booktitle = { 2026 IEEE International Conference on Quantum Software (QSW) },
  title = {{ Software Between Quantum and Machine Learning - and Down to Pulses }},
  year = {2026},
  volume = {},
  ISSN = {},
  pages = {66-78},
  doi = {10.1109/QSW72780.2026.00017},
  url = {https://doi.ieeecomputersociety.org/10.1109/QSW72780.2026.00017},
  publisher = {IEEE Computer Society},
  address = {Los Alamitos, CA, USA},
  month = Jul
}

@InProceedings{strobl_qml_2025,
  author = {Strobl, Melvin and Franz, Maja and Kuehn, Eileen and Mauerer, Wolfgang and Streit, Achim},
  booktitle = {2025 IEEE International Conference on Quantum Software (QSW)},
  title = {{QML}-Essentials -- A Framework for Working with Quantum Fourier Models},
  year = {2025},
  pages = {238--243},
  doi = {10.1109/QSW67625.2025.00035}
}

@article{peters_generalization_2023,
  author = {Peters, Evan and Schuld, Maria},
  journal = {Quantum},
  volume = {7},
  pages = {1210},
  title = {Generalization despite overfitting in quantum machine learning models},
  doi = {10.22331/q-2023-12-20-1210},
  eprint = {2209.05523},
  eprinttype = {arxiv},
  archiveprefix = {arXiv},
  month = dec,
  year = {2023}
}

@article{barthe_gradients_2024,
  title = {Gradients and Frequency Profiles of Quantum Re-Uploading Models},
  author = {Barthe, Alice and {P{\'e}rez-Salinas}, Adri{\'a}n},
  year = 2024,
  journal = {Quantum},
  volume = {8},
  pages = {1523},
  doi = {10.22331/q-2024-11-14-1523},
  eprint = {2311.10822},
  archiveprefix = {arXiv},
  primaryclass = {quant-ph}
}

@book{tao_additive_2006,
  title = {Additive Combinatorics},
  author = {Tao, Terence and Vu, Van H.},
  year = 2006,
  series = {Cambridge Studies in Advanced Mathematics},
  volume = {105},
  publisher = {Cambridge University Press},
  address = {Cambridge},
  doi = {10.1017/CBO9780511755149}
}

@book{meckes_random_2019,
  title = {The Random Matrix Theory of the Classical Compact Groups},
  author = {Meckes, Elizabeth S.},
  year = {2019},
  series = {Cambridge Tracts in Mathematics},
  volume = {218},
  publisher = {Cambridge University Press},
  doi = {10.1017/9781108303453}
}

@misc{ohno_observable_2026,
  title = {Observable-Guided Generator Selection for Improving Trainability in Quantum Machine Learning with a {$\mathfrak{g}$}-Purity Interpretation under Restricted Settings},
  author = {Ohno, Hiroshi},
  year = 2026,
  eprint = {2604.15693},
  archiveprefix = {arXiv},
  primaryclass = {quant-ph},
  doi = {10.48550/arXiv.2604.15693}
}

@article{strobl_fourier_2026,
  doi = {10.1088/2058-9565/ae8df8},
  url = {https://doi.org/10.1088/2058-9565/ae8df8},
  year = {2026},
  month = {aug},
  publisher = {IOP Publishing},
  volume = {11},
  number = {4},
  pages = {045008},
  author = {Strobl, Melvin and Sahin, M Emre and van der Horst, Lucas and Kühn, Eileen and Streit, Achim and Jaderberg, Ben},
  title = {Fourier fingerprints of ansatzes in quantum machine learning},
  journal = {Quantum Science and Technology}
}

@article{brandao_local_2016,
  title = {Local Random Quantum Circuits Are Approximate Polynomial-Designs},
  author = {Brand{\~a}o, Fernando G. S. L. and Harrow, Aram W. and Horodecki, Micha{\l}},
  year = 2016,
  journal = {Communications in Mathematical Physics},
  volume = {346},
  number = {2},
  pages = {397--434},
  doi = {10.1007/s00220-016-2706-8},
  eprint = {1208.0692},
  archiveprefix = {arXiv},
  primaryclass = {quant-ph}
}

@article{arrasmith_equivalence_2022,
  title = {Equivalence of Quantum Barren Plateaus to Cost Concentration and Narrow Gorges},
  author = {Arrasmith, Andrew and Holmes, Zo{\"e} and Cerezo, M. and Coles, Patrick J.},
  year = 2022,
  journal = {Quantum Science and Technology},
  volume = {7},
  number = {4},
  pages = {045015},
  publisher = {IOP Publishing},
  doi = {10.1088/2058-9565/ac7d06}
}
\glsresetall

\appendices

\section{Circuit Ansatzes}

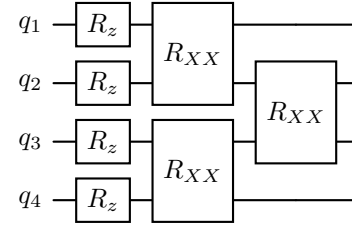
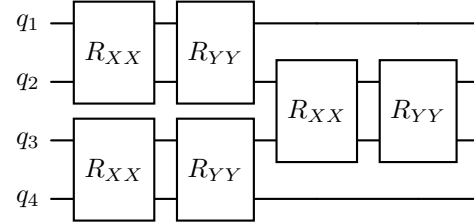
\begin{figure}[!htb]
  \centering
  \begin{subfigure}[b]{\linewidth}
    \centering
    \adjustbox{max width=\linewidth}{%
      \begin{quantikz}[column sep=0.3cm, row sep=0.2cm]
      \lstick{$q_1$} & \gate{\RZ} & \gate[2]{R_{XX}} &                  & \\
      \lstick{$q_2$} & \gate{\RZ} &                  & \gate[2]{R_{XX}} & \\
      \lstick{$q_3$} & \gate{\RZ} & \gate[2]{R_{XX}} &                  & \\
      \lstick{$q_4$} & \gate{\RZ} &                  &                  &
      \end{quantikz}}
    \caption{\Gls{mga}: \gls{dla} $\mso(2n)$, $\dim\mg=n(2n-1)$.}
    \label{fig:mga}
  \end{subfigure}

  \begin{subfigure}[b]{\linewidth}
    \centering
    \adjustbox{max width=\linewidth}{%
      \begin{quantikz}[column sep=0.3cm, row sep=0.2cm]
      \lstick{$q_1$} & \gate[2]{R_{XX}} & \gate[2]{R_{YY}} &                  &                  & \\
      \lstick{$q_2$} &                  &                  & \gate[2]{R_{XX}} & \gate[2]{R_{YY}} & \\
      \lstick{$q_3$} & \gate[2]{R_{XX}} & \gate[2]{R_{YY}} &                  &                  & \\
      \lstick{$q_4$} &                  &                  &                  &                  &
      \end{quantikz}}
    \caption{Off-diagonal ansatz: \gls{dla} $\god\cong\mso(n)\oplus\mso(n)$, $\dim\god=n(n-1)$.}
    \label{fig:odansatz}
  \end{subfigure}
  \caption{
    One layer of the two baseline \ase, drawn for $n=4$ with an independent trainable angle for every gate.
    (a)~The \gls{mga} applies $\RZ$ on every qubit and nearest-neighbor $R_{XX}$ gates.
    (b)~The off-diagonal ansatz replaces the single-qubit rotations by nearest-neighbor $R_{YY}$ gates, generating the floor-free algebra $\god\cong\mso(n)\oplus\mso(n)$.
    The training circuit of~\autoref{fig:precondition_training} uses $11$ such trainable blocks alternating with $10$ $\RY$ encoding blocks, beginning and ending with a trainable block.
  }
  \label{fig:ansaetze}
\end{figure}

\begin{figure}[!htb]
  \centering
  \newcommand{\hardgraph}[2]{%
    \begin{tikzpicture}[x=0.6cm, y=0.6cm, baseline=-2pt]
      \colorlet{ser}{#1}%
      \tikzset{qd/.style={circle, fill=ser, inner sep=0pt, minimum size=4.2pt},
               add/.style={draw=ser, line width=1.1pt}}%
      \foreach \i [evaluate=\i as \q using int(\i+1)] in {0,...,4} {
        \node[qd] (n\i) at (\i,0) {};
        \node[font=\tiny, text=black!55] at (\i,-0.5) {\q};
      }
      \foreach \i [evaluate=\i as \j using int(\i+1)] in {0,...,3}
        \draw[draw=black!55, line width=0.5pt] (n\i)--(n\j);
      #2
    \end{tikzpicture}}
  \begin{subfigure}[b]{0.48\linewidth}
    \centering
    \hardgraph{color3}{}
    \caption{Off-diagonal chain: path, $d_Z{=}0$.}
    \label{fig:hg-chain}
  \end{subfigure}
  \hfill
  \begin{subfigure}[b]{0.48\linewidth}
    \centering
    \hardgraph{color1}{\draw[add] (n1) to[bend left=50] (n4);}
    \caption{$+$ chord $(2,5)$: bipartite, $d_Z{=}0$.}
    \label{fig:hg-chord14}
  \end{subfigure}

  \vspace{0.8em}
  \begin{subfigure}[b]{0.48\linewidth}
    \centering
    \hardgraph{color4}{%
      \draw[add] (n0) to[bend left=55] (n2);
      \draw[add] (n1) to[bend left=55] (n3);
      \draw[add] (n2) to[bend left=55] (n4);}
    \caption{$+$ $X_kY_{k+2}$: doped chain, $d_Z{=}0$.}
    \label{fig:hg-doped}
  \end{subfigure}
  \hfill
  \begin{subfigure}[b]{0.48\linewidth}
    \centering
    \hardgraph{color2}{\draw[add] (n0) to[bend left=60] (n2);}
    \caption{$+$ chord $(1,3)$: odd cycle, $d_Z{>}0$.}
    \label{fig:hg-odd}
  \end{subfigure}
  \caption{
    Interaction graphs of the families in~\autoref{fig:hollow}, drawn for $n=5$ with qubits $1,\dots,5$ and color-matched to that figure.
    The path and graph chords carry independently parameterized $R_{XX}$ and $R_{YY}$ gates, so the bare path (a) is the off-diagonal \as of~\autoref{fig:odansatz}.
    Adding the chord $(2,5)$ (b) keeps the graph bipartite, and the doped chain (c) adds the next-nearest-neighbor $X_kY_{k+2}$ bonds.
    The colored dopant bonds in (c) instead carry $R_{XY}$ gates.
    Both perturbations enlarge $\dim\mg$ while keeping $d_Z=0$ at the tested sizes.
    The chord $(1,3)$ (d) closes a triangle, an odd cycle that restores the floor $d_Z>0$.
  }
  \label{fig:hard-graphs}
\end{figure}
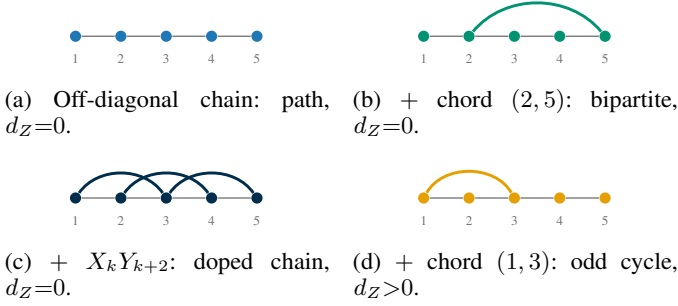

\autoref{fig:ansaetze} shows one layer of each baseline \as.
The variants in~\autoref{sec:results} either add next-nearest-neighbor $R_{XY}$ gates to the off-diagonal circuit or place its $R_{XX}$ and $R_{YY}$ pairs on the edges of a non-path graph.
\autoref{fig:hard-graphs} shows their interaction graphs.

\section{Purity Convention and Extremal Inputs}
\label{apx:purity}

\paragraph{Normalization}

Each bare Pauli string differs from its orthonormalized counterpart in~\citeauthor{ragone_lie_2024}~\cite{ragone_lie_2024} by a factor $2^{\nicefrac{n}{2}}$.
Consequently, the purities satisfy $\pur_\mg(\rho)=2^n\pur^{\mathrm{HS}}_\mg(\rho)$.
For a Pauli-string observable $\obs\in \imag\mg$, we also have $\pur^{\mathrm{HS}}_\mg(\obs)=\tr[\obs^2]=2^n$.
These factors cancel in~\autoref{eq:variance}, leaving $\var_{\btheta}[f_{\btheta}]=\pur_\mg(\rho)/\dim\mg$ for a simple algebra.
For $n=3$ or $n\ge5$, the readout $\obs=X_iX_{i+1}+Y_iY_{i+1}$ contributes one basis string to each of the two equal-dimensional simple ideals $\mg_1,\mg_2$ of $\god$~\cite{wiersema_classification_2024}.
With input purities $\pur_{\mg_1}+\pur_{\mg_2}=\pur_\god$, \autoref{eq:variance} gives $\var_{\btheta}[f_{\btheta}]=2(\pur_{\mg_1}+\pur_{\mg_2})/\dim\god=2\pur_\god/\dim\god$.
Smaller systems, however, require separate accounting.
At $n=4$, each $\mso(4)$ splits into two simple ideals, each of dimension $3$ and observable \gls{hs} purity $8$, so the same aggregate factor holds.
At $n=2$, both components are abelian and the output variance is zero.

\paragraph{Extremal Inputs}

For $n\ge2$, the clustered value $n-1$ reached at $\phi_k=\nicefrac{\pi}{2}$ is the global maximum of~\autoref{eq:offdiag-closedform}.
To see this, write $s_k=\sin^2\phi_k$ and $t_k=\cos^2\phi_k$ and collect the terms of~\autoref{eq:offdiag-closedform} with right endpoint $k$ into $A_k=\sum_{j<k,\,k-j\ \mathrm{odd}}\, s_j\prod_{j<l<k}t_l$, together with its even-distance counterpart $B_k$.
These sums obey the recursions $A_{k+1}=s_k+t_kB_k$ and $B_{k+1}=t_kA_k$ with $A_2=s_1$ and $B_2=0$, so $A_k,B_k\le1$ follows by induction.
Since the purity accumulates as $\pur_\god=\sum_{k=2}^{n}s_kA_k$, this yields $\pur_\god\le n-1$, with equality exactly when $\phi_k\in\{\nicefrac{\pi}{2},\nicefrac{3\pi}{2}\}$ at every site.
A randomized search over angle configurations up to $n=8$ found no violation of this bound.

\section{Input Transformation}
\label{apx:input_transformation}

\paragraph{Acceptance Test}

For $n\ge3$, let $\mathcal{D}$ be the empirical distribution of $m$ fixed, nonzero inputs in $\mathbb{R}^D$, where $D\ge2n$.
Draw a shared rotation $Q$ from the Haar measure on $SO(D)$.
For each fixed input, $Q\bx/\lVert\bx\rVert$ is uniform on the unit sphere and has the same distribution as $\boldsymbol{z}/\lVert\boldsymbol{z}\rVert$ for $\boldsymbol{z}\sim\mathcal{N}(0,I_D)$~\cite[Sec.~1.2]{meckes_random_2019}.
This common normalization does not affect polar angles, and disjoint Gaussian coordinate pairs have independent uniform angles by rotational invariance~\cite[Sec.~1.2]{meckes_random_2019}.
Each pair has nonzero radius almost surely, so the angle map is well defined on the finite dataset with probability one.
Although the same $Q$ is used for all inputs, linearity of expectation~\cite[Chap.~1]{tao_additive_2006} then gives
\begin{equation}
  \mathbb{E}_Q\!\left[\, \mathbb{E}_{\bx\sim\mathcal{D}}\!\left[ \pur_\god\!\left(\rho(\bphi(Q\bx))\right) \right] \right] = \mu_n,
  \label{eq:fixedq-mean}
\end{equation}
where $\mu_n$ is the uniform-prior mean in~\autoref{eq:offdiag-mean}.
The dataset mean lies in $[0,n-1]$ by~\autoref{apx:purity}.
Applying Markov's inequality~\cite[Chap.~1]{tao_additive_2006} to $n-1-\mathbb{E}_{\bx}[\pur_\god]$ gives
\begin{equation}
  \Pr_Q\!\left[\, \mathbb{E}_{\bx\sim\mathcal{D}}\!\left[ \pur_\god \right] \ge \frac{\mu_n}{2} \right] \ge \frac{\mu_n/2}{n-1-\mu_n/2} \xrightarrow{\;n\to\infty\;} \frac{1}{5} .
  \label{eq:fixedq-prob}
\end{equation}
Because~\autoref{eq:offdiag-closedform} gives empirical mean purity, we can recognize an accepted draw without simulating the circuit.
For a dataset $\{\bx_1,\dots,\bx_m\}$ and candidate $Q$, that quantity is
\begin{equation}
  \hat{\pur} \;=\; \frac{1}{m}\sum_{i=1}^{m} \pur_\god\big(\rho(\bphi(Q\bx_i))\big),
  \label{eq:uniformity-test}
\end{equation}
which can be evaluated from~\autoref{eq:offdiag-closedform} in $\mathcal{O}(mn^2)$ classical time, or in $\mathcal{O}(mn)$ using the recursions in~\autoref{apx:purity}, once the angles are available.
We accept a rotation when $\hat{\pur}\ge\mu_n/2$, the event whose probability is bounded in~\autoref{eq:fixedq-prob}.
Independent rejection sampling then needs at most $2(n-1)/\mu_n-1$ draws in expectation, with the bound approaching five as $n\to\infty$.
Each trial also incurs the cost of generating and applying $Q$.

\begin{figure}[tb]
  \centering
  \input{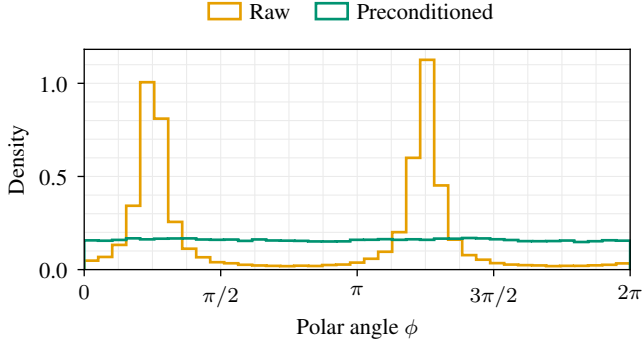}
  \caption{Polar angles of correlated coordinate pairs with Pearson correlation $0.97$, pooled across samples and sites, before and after a shared random rotation.
    The raw angles cluster near $\pi/4$ and $5\pi/4$.
    The transformed pooled histogram is approximately uniform on $[0,2\pi)$.}
  \label{fig:polar}
\end{figure}

For the constructed correlated datasets in~\autoref{fig:polar}, each containing $m=256$ inputs, all $10^3$ sampled rotations passed at $n=6$, $10$, and $14$.
Their mean $\mathbb{E}_Q[\hat{\pur}]$ agreed with $\mu_n$ to within one percent, although these observations do not determine a worst-case acceptance rate.
Indeed, this test measures purity rather than uniformity: the raw angles in~\autoref{fig:polar}, clustered near $\nicefrac{\pi}{4}$, can also pass without rotation.
By comparison, angles clustered near $\{0,\pi\}$ lie about two orders of magnitude below the threshold and are rejected.

For the specified readout and circuit ensemble, every accepted rotation certifies empirical mean sample-wise output variance $\Theta(\nicefrac{1}{n})$.
It does not control the variance of an averaged output or a supervised-loss gradient, which also involve cross-sample covariances.
Selecting $Q$ on a finite dataset likewise gives no population guarantee without further analysis.

\paragraph{Effect on the Angle Distribution}

The pooled angle histogram in~\autoref{fig:polar} becomes flatter after the shared rotation and polar-angle extraction.
Because the histogram pools sites, it does not establish uniformity at each site or independence between them.
Instead, the acceptance test evaluates empirical mean purity directly, without requiring either property.

\paragraph{Effect of Weighted Encodings}

Across integer-weight scalar encodings, the exact clustered endpoint behaves consistently: $x\in\{0,\pi\}$ sends every angle to $\{0,\pi\}$.
\autoref{eq:exact-zero} therefore applies to the floor-free models at these inputs, including with re-uploading.
For a single off-diagonal encoding block, \autoref{eq:offdiag-closedform} has additional zeros when all but one qubit clusters, as at $x=\nicefrac{\pi}{2}$ with binary weights or $x=\nicefrac{\pi}{3}$ with ternary weights.

Away from exact clustering, the weights change how strongly scalar-input jitter reaches each site.
In~\autoref{fig:encodings-mean}, exponential weights amplify a small jitter at high-weight sites and partially recover mean purity, while equal weights remain strongly suppressed.

\begin{figure}[tb]
  \centering
  \input{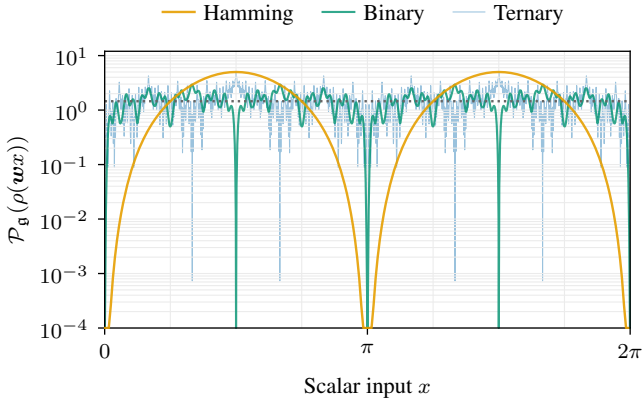}
  \caption{Input-purity landscapes $\pur_\god(\rho(\bw x))$ for the $\RY$ encoding at $n=6$.
    Equal, binary, and ternary weights have allowed frequency-set sizes $|\Omega|=13$, $127$, and $729$, respectively, and the dotted line marks the \iid uniform mean.
    Binary and ternary weights produce rapid oscillations whose uniform-scalar means equal the \iid mean (\autoref{eq:spectral-precondition}).
    The binary curve has exact zeros at $x\in\{0,\nicefrac{\pi}{2},\pi,\nicefrac{3\pi}{2}\}$.
    These sizes count allowed model frequencies in $\Omega$.
    The plotted purities have frequencies in $2\Omega$, with some coefficients equal to zero.}
  \label{fig:encodings-landscape}
\end{figure}

At $n=14$ and $\sigma=0.03$, the clustered mean for equal weights lies $4.6$ orders of magnitude below its uniform value, whereas the binary-weight mean is within a factor of two.
These results describe the tested scalar priors rather than guaranteeing large purity for every distribution with a large allowed spectrum.
Across these weighted encodings, the algebra and computational-basis clustering point are unchanged, so their diagonal count remains the same.
Away from that endpoint, the landscapes in~\autoref{fig:encodings-landscape} show how induced angles affect purity.

\begin{figure}[tb]
  \centering
  \input{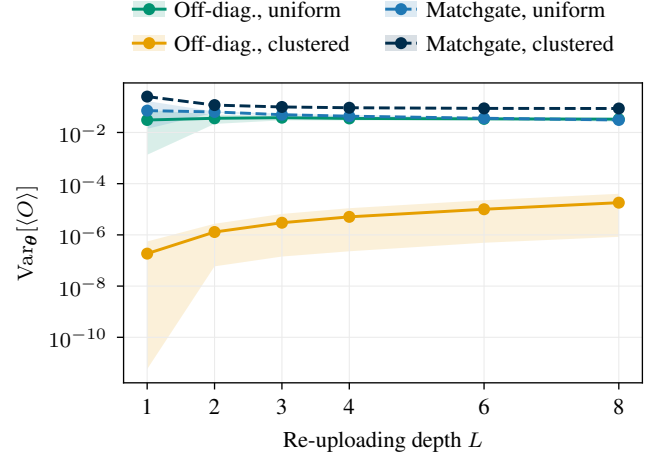}
  \caption{Re-uploading depth at $n=6$, with $10^3$ independent uniform parameter draws from $[0,2\pi)$ per input.
    Curves show means and bands the $10$--$90\%$ range over $10$ angle configurations.
    Here the clustered prior uses independent $\phi_k\sim\mathcal{N}(0,0.03^2)$.
    The off-diagonal readout is $X_iX_{i+1}+Y_iY_{i+1}$ (solid), and the matchgate control uses $Z_i$ (dashed).
    At $\sigma=0$, the off-diagonal output vanishes at every depth (\autoref{eq:exact-zero}).}
  \label{fig:reupload}
\end{figure}

\newpage
Finally, we compare re-uploading depths in~\autoref{fig:reupload} using the same ansatz-first convention as the training experiment.
Each circuit has $L$ encoding blocks and $L+1$ trainable blocks.
The exact zero-output statement holds at every depth for computational-basis inputs, while the uniform-angle branch has no comparable multi-block closed form.
At $n=6$, its measured mean sample-wise output variance lies between approximately $0.031$ and $0.038$ for $L\in\{1,2,3,4,6,8\}$.
This range describes stability relative to the observed $L=1$ circuit and is not an equality with the single-block Haar prediction.
For clustered angles with $\sigma=0.03$, the variance remains three to five orders of magnitude lower and rises slowly with depth.
At $\sigma=0$, the measured output satisfies $\max|\braket{\obs}|<10^{-13}$ at $L=1$ and $L=8$, consistent with~\autoref{eq:exact-zero}.
The matchgate control retains nonzero variance under both priors, although the values differ by factors of approximately $2$ to $3.5$ over the measured depths.

\end{document}